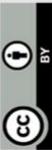

Original Article

# Astro-Animation - How Artists and Scientists Envision the Universe

Laurence ARCADIAS [1,*], Robin H.D. CORBET [2] and Emma BOOTH [3]

[1] Maryland Institute College of Art; Liverpool John Moores University; Transart Institute; larcadias@mica.edu. ORCID ID: 0000-0002-5062-0238
[2] University of Maryland, Baltimore County, NASA Goddard Space Flight Center, Maryland Institute College of Art; corbet@umbc.edu ORCID ID: 0000-0002-3396-651X
[3] Maryland Institute College of Art, University of Maryland, Baltimore County, NASA Goddard Space Flight Center; emmabooth14@gmail.com
[*] Correspondence: larcadias@mica.edu

**Abstract:** For several years, students at an art college, working with NASA astronomers, have produced animations inspired by research on black holes, dark matter and more. They can be whimsical or poetic but still constrained by scientific rigour. The animations are used for scientific outreach and are freely available. Our program received a positive assessment through an evaluation we undertook. We are now planning a mobile STEAM exhibition to engage teenagers from underrepresented communities who may not typically consider STE(A)M for their studies. "Science anxiety" has been reported to be a significant barrier to learning. Mixing animation with astronomy can stimulate interest in STEAM, making science engaging in an unconventional way. One component would be activities where participants create artistic responses to astronomy. We undertook a workshop at a local city-run school, specialising in the arts for ages 14-17, to brainstorm the art/science activities. There we gave short scientific presentations leading to art activities: a giant colouring wall with projected celestial phenomena, a stop-motion station, and colouring images of comet 67P to produce an animation. Surveys before and after the activities showed positive responses. The hand of the artist has long been an important concept in animation (Crafton 1991). In a film entitled "The Movements of the Universe", this concept is adapted to the hands of scientists. Combining animation, filmed interviews at NASA (including a Nobel prize winner), and the scientists' hands, bring unexpected feelings of dream and humour to the audience. In this paper we explore three different viewpoints of these activities from a scientist, an animator, and an animation student.

## 1. Introduction - A Comparison of Views

This paper explores how art and science can come together to make the invisible universe visible—and emotionally accessible—through collaborative animation. We examine this through three distinct lenses: that of a scientist, an animator and professor of animation, and an animation student who interned at NASA. Each offers a unique perspective on how complex astronomical phenomena can be interpreted and communicated through visual storytelling.

This paper begins with a review of the theoretical frameworks that underpin visual science communication and interdisciplinary collaboration. Next, we introduce the Astro-animation project and its core components. We then present three perspectives—those of a scientist, an animator, and a student—to illustrate how Astro-animation operates in practice. This is followed by an analysis of educational and public engagement outcomes. Finally, we discuss the broader implications for science communication and STEAM education.

## 2. Literature Review / Background Theoretical Framework: Narrative, Inclusion, and Cultural Framing in Science Communication

Astro-Animation's integration of storytelling with astronomical content aligns with a growing body of research that positions narrative as a powerful tool for science communication. Jerome Bruner (1986) distinguishes between two fundamental modes of human understanding: the logico-scientific, which relies on empirical reasoning and structured explanation, and the narrative, which constructs meaning through story, emotion, and context. Avraamidou and Osborne (2009) build on this distinction, arguing that blending these modes enhances accessibility and engagement in public science contexts—a strategy central to Astro-Animation's student films and public workshops. Egan (1986) similarly views storytelling as a primary pedagogical structure, supporting the use of narrative arcs in educational design to deepen learners' emotional and conceptual connections with scientific content.

This narrative framing is also a tool for inclusion. Archer et al.'s (2015) theory of science capital shows how social and cultural resources shape access to science participation. Astro-Animation expands this capital by offering culturally relevant, emotionally engaging ways into STEAM for underrepresented youth. This is especially significant in astronomy, a subject shown to inspire cross-disciplinary curiosity but often underrepresented in formal education (Salimpour et al., 2021).

Finally, Salimpour and Fitzgerald (2024) argue that astronomy is not only scientific but deeply cultural, understood through symbolic, narrative, and embodied lenses across societies. Astro-Animation's use of gesture, metaphor, and visual storytelling supports this pluralistic vision, reframing scientific communication as a form of semiotic translation rather than a simplification.

These theoretical perspectives lay the groundwork for the interdisciplinary practices explored in the next sections. We begin with the scientist's view, examining how astronomical knowledge is constructed, visualised, and shared—and how collaboration with animators reveals new layers of meaning in both research and communication.

**The Astro-Animation Project: An Overview**

The Astro-animation initiative began with a course at the Maryland Institute College of Art, co-taught by an animator and a NASA astrophysicist. Students work with scientists to translate current research topics—such as black holes, pulsars, and lunar missions—into short, animated films. The course integrates scientific accuracy with metaphorical and poetic expression.

Over time, the project expanded to include public workshops and a traveling exhibition titled *Look Up at the Sky, Draw Down the Stars*. These workshops take place in libraries, conferences, and schools, particularly targeting underserved youth. Participants learn basic astronomical concepts through short talks, then create drawings or animations that interpret these ideas. These experiences culminate in collaborative films that are shown at festivals and exhibitions.

The project also developed a toolkit to support facilitators, including step-by-step instructions, visual aids, and evaluation materials.

**4. Perspectives on Astro-Animation**

### 4.1  How a Scientist Envisions the Universe

Astronomers follow the general scientific method of performing observations, formulating hypotheses, testing these, and then reformulating hypotheses if required and conducting further tests. Astronomy, however, differs from many other fields of science in that typically direct experiments cannot be performed, but instead "natural" experiments must be found in the Universe. Also, the scope of astronomy is incredibly broad, comprising the entire Universe since its birth, and can involve physics in much more extreme conditions of temperature, gravity, density and magnetic fields than can exist on the Earth. Astronomers' views of the Universe primarily come from light, although the range of light that be studied has increased, and other "messengers" can now also be used such as cosmic rays, neutrinos, and gravitational waves. In the next section I describe the various types of light, with an emphasis on the high-energy regimes of X-rays and gamma-rays, which I am primarily involved with, and give an example of a type of binary star system that emits at these wavelengths.

**The Different Types of Light**

The light that we can see with our own eyes is only a small slice of the broad spectrum of light that exists - or electromagnetic spectrum to use the technical term. Beyond the reddest light that the human eye can see exists infrared light, going beyond this to redder and redder parts of the spectrum, we reach radio waves of longer and longer wavelengths. In the other direction beyond the bluest light, we can directly perceive exists ultraviolet light, going further we find X-rays and then gamma rays. We note in passing that the centuries long debate about whether light is a wave or a particle was resolved by quantum mechanics which shows that light has both properties.

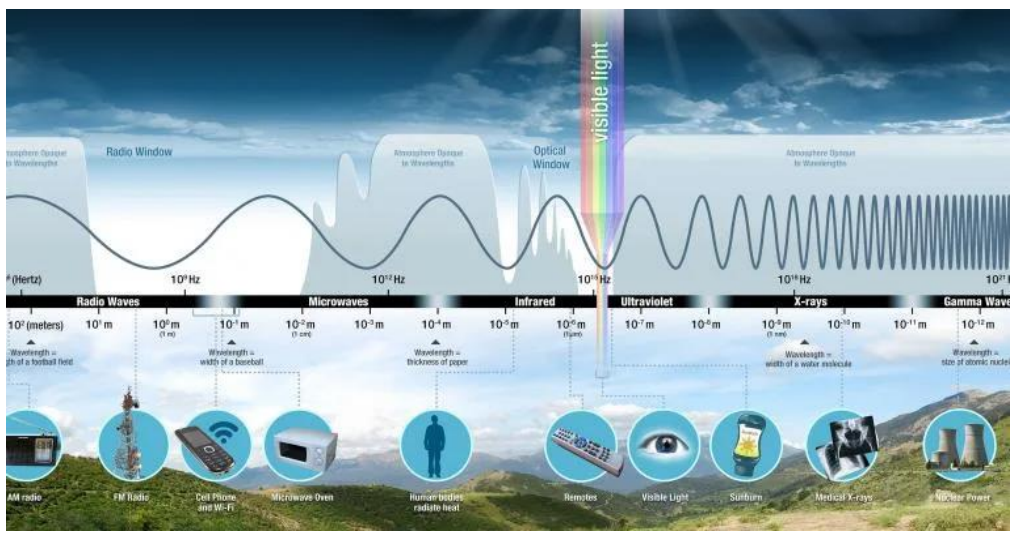

**Fig. 1.** The electromagnetic spectrum, covering, from left to right, radio waves to gamma rays. The diagram indicates wavelength and frequency and also shows atmospheric transparency at each wavelength. Visible light occupies a narrow band, highlighting the limited range perceptible to the human eye. At the bottom are shown examples of terrestrial sources of "light" at each wavelength. (Courtesy NASA)..

The Fermi Gamma-ray Space Telescope was launched into low-Earth orbit in 2008. The mission is an international collaboration led by NASA. It carries two instruments to detect gamma rays, which are the most energetic form of "light". The main instrument onboard Fermi is the Large Area Telescope or LAT (Atwood et al., 2009). While the telescope name is not very exciting, the LAT studies some of the most extreme and exotic conditions in the Universe, such as the environments around black holes, or rapidly rotating neutron stars, the remains of a massive star after it has exploded as a supernova.

The mirrors and lenses which are used for telescopes that detect visible light do not work for gamma rays. The LAT instead relies on the creation of antimatter within the telescope (e.g. Funk, 2015). When antimatter and matter meet they annihilate each other and produce a pair of gamma-ray photons. Thanks to Einstein's $E = mc^2$, the process can also work in the other direction and a gamma ray, in the right situation, can convert into antimatter and matter. In the LAT a gamma ray can hit an atom in a metal sheet and change into an electron, and the antimatter version of an electron known as a positron. As the electron and positron move through sheets of silicon they produce an electronic signal that is then transmitted to the ground. Subsequently computer algorithms calculate where on the sky the gamma came from, when it arrived, and how much energy it had.

This computer file, which contains a list of information about each detected gamma ray, is what scientists use to study the very high-energy Universe. While the LAT does not produce a direct image of the Universe within it, this computer file can be used to produce a gamma-ray view of the sky.

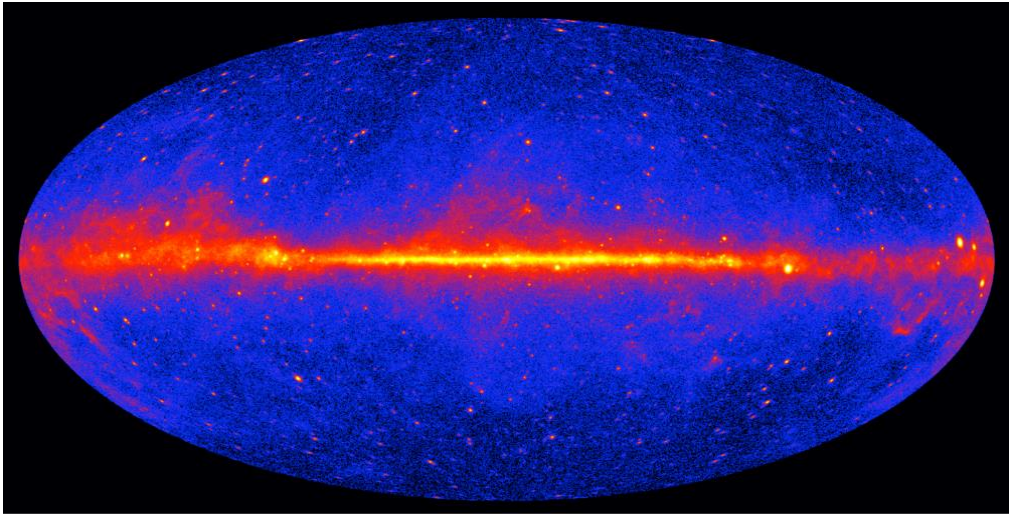

**Fig. 2.** The entire gamma-ray sky as viewed by the LAT detector on board the Fermi satellite. The centre of the image corresponds to the centre of the Milky Way Galaxy. (Courtesy NASA).

### 3.2 "**Carnivorous Spider**" **Binary Star Systems**

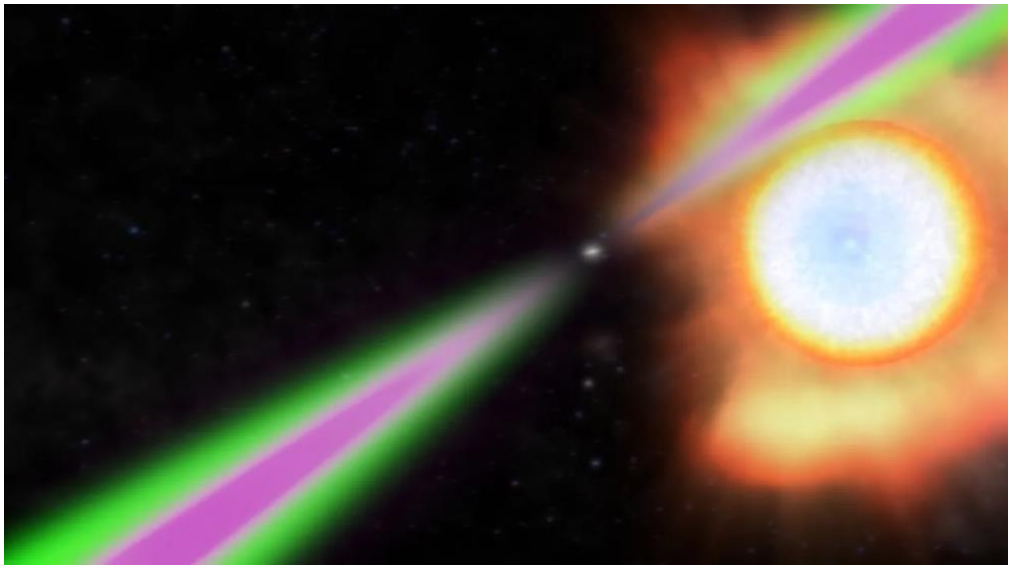

**Fig. 3.** An artist's concept of a black-widow binary star system produced by the NASA Goddard Scientific Vizualization Studio. (Courtesy NASA).

An example of the type of object that the Fermi LAT studies is a special type of binary star system. In a binary star system the two stars orbit around each other, similarly to how the Earth and the Moon orbit each other. In this particular system one of the components is a rapidly rotating neutron star, a pulsar. A neutron star is the core of a massive star left after a supernova explosion that has been compressed down to the size of a large city but has a mass of approximately 1.4 times that of the Sun (e.g. Shapiro & Teukolsky, 1983). In this object most electrons and protons are forced to combine together into neutrons, and it is essentially a giant atomic nucleus with incredibly strong gravity and magnetic field. Many of these objects are pulsars that emit regular flashes of electromagnetic radiation. In some systems the neutron star can rotate very rapidly, taking just a few milliseconds to complete one rotation. The energy emitted from the pulsar can be enough to slowly evaporate its companion. These have been named after two species of carnivorous spiders - black widows and redbacks (e.g. Roberts, 2013).

The question then arises of how to communicate the excitement scientists feel for a subject such as the "spider binaries" with a general audience, and how best to deal with the physics involved. Traditionally NASA has provided films that attempt to be rather strict visualisations of a subject. But this is often impossible, and sometimes the underlying sense of wonder may be lost. An alternative approach, and one that is described in this paper, is that while there must be underlying scientific accuracy, it can often be more inspiring to provide animations that take a much freer approach to the material.

**Creation of an Astro-Animation Class**

To explore the interactions between science and art in a way that can present scientific results in a very different way we created an undergraduate class in Astro-Animation at an art college. The class consists of two interlocking parts, one focused on astronomy and one on animation. The class starts with the astronomy section, taught by the author of this section, and this meets the science requirements for the students' Bachelor of Fine Arts degrees. For the animation component, the students are supervised by the author of section two, but they also work directly with NASA scientists as mentors for one semester. The students produce animations on a wide variety of astronomical topics that have included black holes, life on Mars, gravitational waves, and many others. The students are asked to incorporate the scientific concepts in their animations through artistic vision, but still with underlying scientific rigour. For example, they can use metaphorical or poetic interpretations. The students are given an overview of the scientific process including rigorous testing of hypotheses as part of the astronomy component of the class.

**Astro-Animation class timeline**

During the first two weeks, students learn about the overall class goals, and undertake some warmup exercises doing very brief animations on drawings on either the scale of the Universe or gravity. Then, in the third week, the NASA scientist mentors come to the class and give a couple of minute presentations on the topics they are working on. Typically, there are six projects. Teams of three students are then randomly assigned to each topic/scientist, and the animators and scientists discuss their thoughts on how the project will proceed.

For the next two weeks, the student animators work on their ideas and put together short presentations incorporating animatics and sketches. Then they are taken on a visit to the NASA Goddard Space Flight Center. After a tour of the centre, where they can view satellites being constructed and tested, they present their ideas to a group of scientists and science outreach experts. They then again split into groups to receive detailed feedback and discuss how the project will continue for the rest of the semester. After returning to the art school, they work on the animations and get regular feedback from the animation instructor and other students, as well as remaining in close contact, typically remotely, with the scientist mentors.

The students again return to NASA GSFC one week before the end of the semester to screen their films. Coincidentally, the day of the class coincides with the annual Take Your Child to Work Day, and the students have a large audience of both adults and children. Afterwards, a few final tweaks and corrections may be made to the films.

Currently there is a collection of 73 short animations that have been made through our class that are freely available, and this number continues to increase each year. These have been used in a variety of ways by scientists, teachers, and outreach specialists. In addition, the animations have been shown at a variety of venues over the years that have included art and film festivals, science events, and science fiction conventions.

**Evaluation of the Astro-Animation Project**

We received funding from the US National Endowment for the Arts (NEA) to evaluate the benefits and outcomes of our class. We conducted surveys and interviews of scientists, animation students, and non-specialist audiences. From these, we found that presenting astronomy via animation is highly attractive to a range of audiences. Just screening the animations on their own has a visual impact. However, for science learning it is valuable to have some additional scientific material (Arcadias et al., 2020

**Expanding Beyond the Classroom to Unusual Locations - A Pilot Study**

With the success of our class in getting students enthusiastic about astronomy, and the sense of wonder that the animations helped to bring to the audiences for these films, we wanted to build on this to reach out to a much wider community. A particular goal is to bring science and art to underrepresented teenagers outside of a traditional classroom.

Our concept is to construct an exhibition that would combine animations with deeper scientific information in events that would involve participants as active members, and not just passive viewers. The plan is to incrementally develop, test and expand the exhibition concept, taking into account feedback from stakeholders at each step.

The working title for the exhibition is "Look up at the Sky, Draw Down the Stars" that reflects both the astronomical content, together with the participants involvement with art/science activities. The exhibition would involve a variety of collaborators bringing a range of expertise. It would include people at our art school, a large research university with expertise in both scientific research and science education, outreach specialists at NASA, and community leaders at the library.

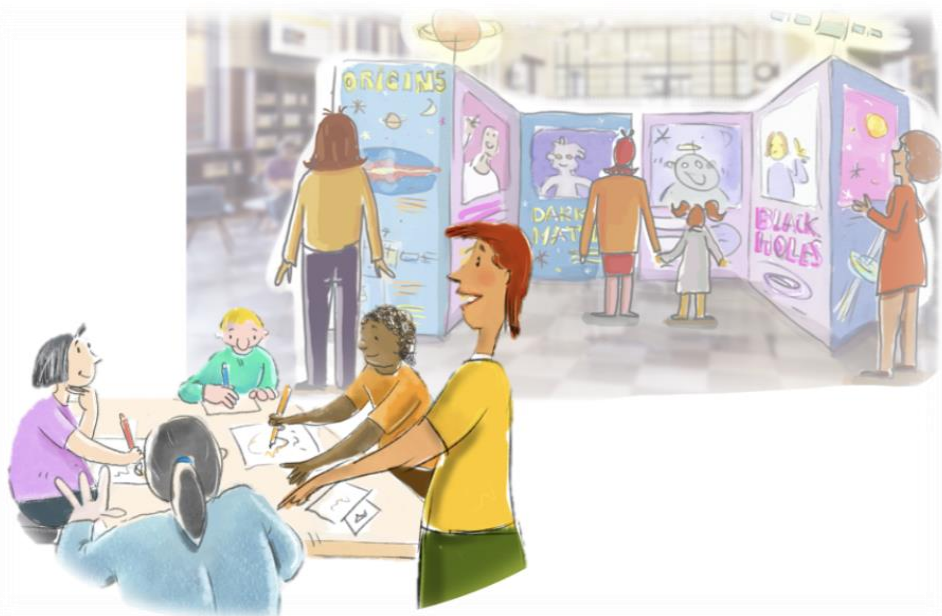

Fig. 4. The concept for the "Look Up at the Sky, Draw Down the Stars" exhibition.

**Astronomy topics, animations, and interviews of scientists**

The exhibition would include several panels, each of which with a particular focus combining a relevant animation and the scientific background. An introductory panel would focus on women astronomers through time and Indigenous artists' views of the sky. For a first location of such an exhibit beyond the art school, the local large public library in our city provides an ideal location. The library provides significant community youth services to attract participants. At the same time, it is a somewhat curated environment that provides physical security for equipment.

A crucial part of the planned exhibition is to include hands-on art and science activities for the audience to make them fully a participant rather than just a viewer. Before moving to a full exhibition at the library, we wanted to prototype the various activities. We therefore conducted workshops at a local high school to test the concepts. This was at a city-run school that is specialised in the arts. Two sessions were held, with school educators also taking part.

Two sessions were held, with educators at the school also contributing to the process. To evaluate the workshops, anonymous surveys were given to the school students before and after each session. These both collected demographic information and students' attitudes toward science and animation to determine how these were affected by the workshops. Six scaled questions on attitude were given before, and eight were given after the activities. The latter survey also gave the students the opportunity to provide free-form responses). For the list of questions asked, see Arcadias & Corbet 2022, figures 4 and 5

**Conclusion to a Scientist's Vision of the Universe**

As a scientist, I find it easy to become lost in my day-to-day of creating graphs of scientific values and applying equations to determine the details of an astronomical phenomenon I am working on. Participating in the astro-animation project has enabled me to step back a little and see more of the overview. In describing fundamental astronomical principles and how my work fits into this context has also provided me with perspective. It has been very satisfying to see the great responses to combining animation with astronomy. An itinerant exhibition could make astronomy and animation available to a broad audience, especially teenagers. We believe this can help to increase diversity in STEAM.

### 4.2 Astro-Animation, How an Artist Envisions the Universe

While the first part of this research focused on interdisciplinary collaborations to engage students and underrepresented communities in STEAM education, this section delves deeper into the creative process of astro-animation. As an artist and professor of animation, I have developed techniques to visually interpret complex astronomical concepts through animation. By integrating artistic methods—such as hand gestures and visual metaphors—I aim to make the invisible aspects of the universe both accessible and engaging to the public.

**The Movements of the Universe: The Role of Hand Gestures**

By applying these principles of combining art and science, I created a film called "*The Movements of the Universe"* that further explores how to represent graphically complex astronomical ideas. The film uses the scientists' hand gestures enhanced by animated visual metaphors, to communicate intricate scientific concepts. I observed that these hand gestures often extend the scientists' thoughts, serving as a bridge between abstract ideas and their physical representation. This inspired me to incorporate hand gestures as a key element in my animation process. The insights I gained while working with NASA scientists during *The Movements of the Universe* motivated me to further explore the role of hand gestures in scientific communication (Roth & Lawless, 2002).

**Analysing Hand Gestures in Astronomy**

In the Astro-Animation class, which I co-teach with a NASA astronomer, I noticed that many scientists naturally use hand gestures to convey complex ideas to students. These gestures visually illustrate abstract phenomena, such as the movement of galaxies or the expansion of the universe, and have sparked my interest in capturing, analysing, and expressing these movements through animation.

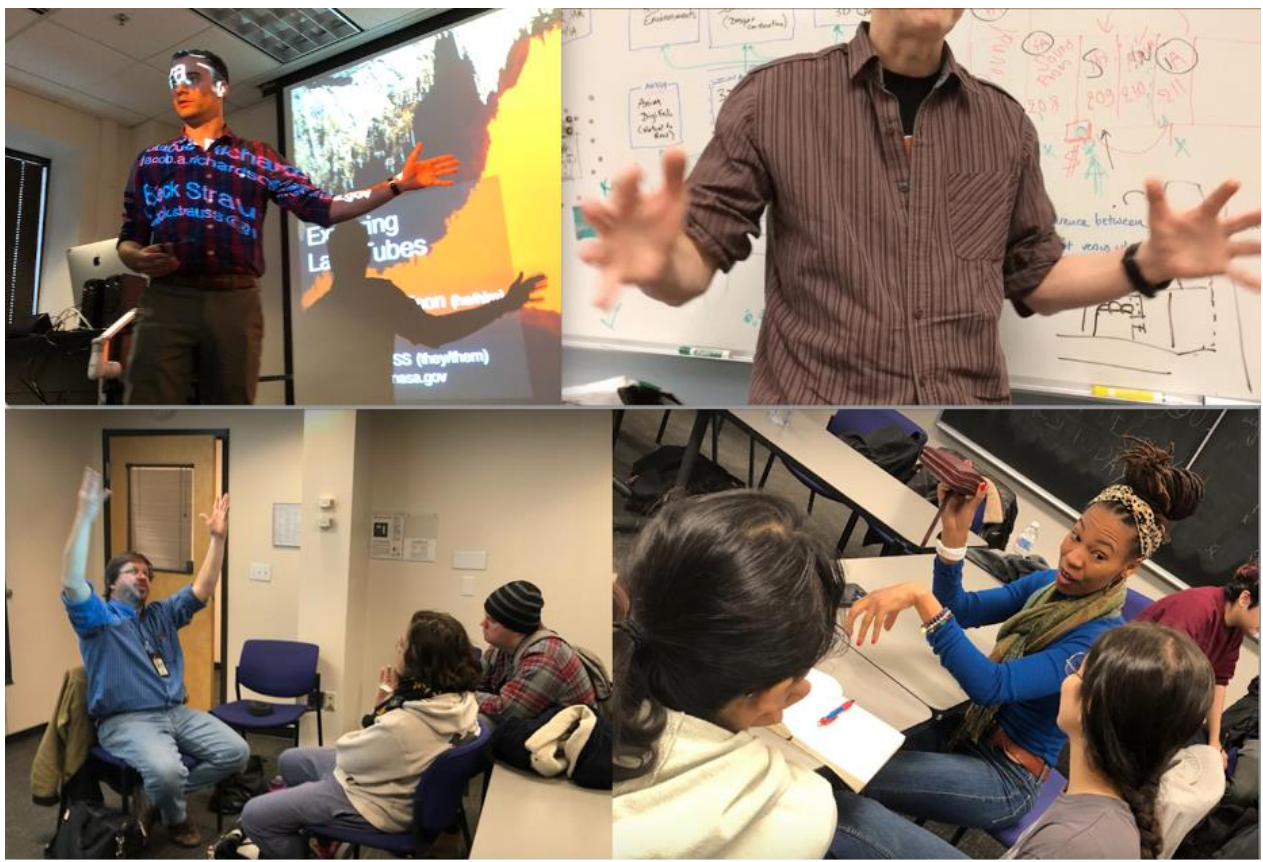

**Fig. 5.** Scientists using expressive hand gestures to explain scientific concepts to students.

This process was deeply influenced by Donald Crafton's concept of the 'Hand of the Artist, outlined in his essay on early animation (Crafton, 1991). Crafton describes how, in the early days of animation, the animator's hand served as a bridge between the audience and the artwork. The animator's hand was a tool to signal the "special magical properties" of the animated film. By showing the animator at work, the hands became the mediator between the spectator and the drawings that 'come to life'. Similarly, in my work, I used the scientists' hand gestures to connect abstract scientific ideas with the audience, transforming these invisible forces of the universe into tangible, visual experiences. This approach informed many of the creative decisions I made during the production of the film.

**The Movements of the Universe – Process and Insights**

Building on my initial observations of scientists' hand gestures, I sought to further explore how these physical movements could serve as visual metaphors in the animation process. I began by sketching ideas on how various astronomical phenomena move, letting my curiosity guide the process. Whenever a particular concept intrigued me, I attempted to make sense of it through these sketches.

To bring these ideas to life, I collaborated with NASA scientists, focusing on their use of hand gestures to explain complex topics such as the Big Bang, the Sun, dark matter, pulsars, gravitational waves, black holes, and binary stars. By observing their gestures, I could capture the essence of their explanations and translate these physical movements into visual metaphors through animation.

**Case Studies: The Role of Hand Gestures in Astro-Animation**

The interviews with the seven scientists provided invaluable insights into how their hand gestures convey complex ideas. From these observations, I was able to develop animations that visually captured these physical movements, transforming abstract concepts into tangible visual metaphors. Two case studies from *The Movements of the Universe* highlight how hand gestures can serve as powerful tools for visual storytelling.

**Case Study 1: The Black Hole Dance**

In my animation film, one scientist explained black holes, describing how they spin and wobble around each other, likening the motion to a dance. His hands moved in a fluid, almost playful way, which I represented in the animation as two juggling hands with black holes spinning overhead. His feedback highlighted that the animation mirrored the mathematical shift of black holes from theoretical to observable entities, bridging the gap between abstract theory and physical observation. Hand gestures were able to communicate the speaker's excitement about their topic and it may help memorise concepts. He also mentioned that he does not usually use his hands when he talks, as it is not part of his culture, and he had to integrate that later into his communication tools.

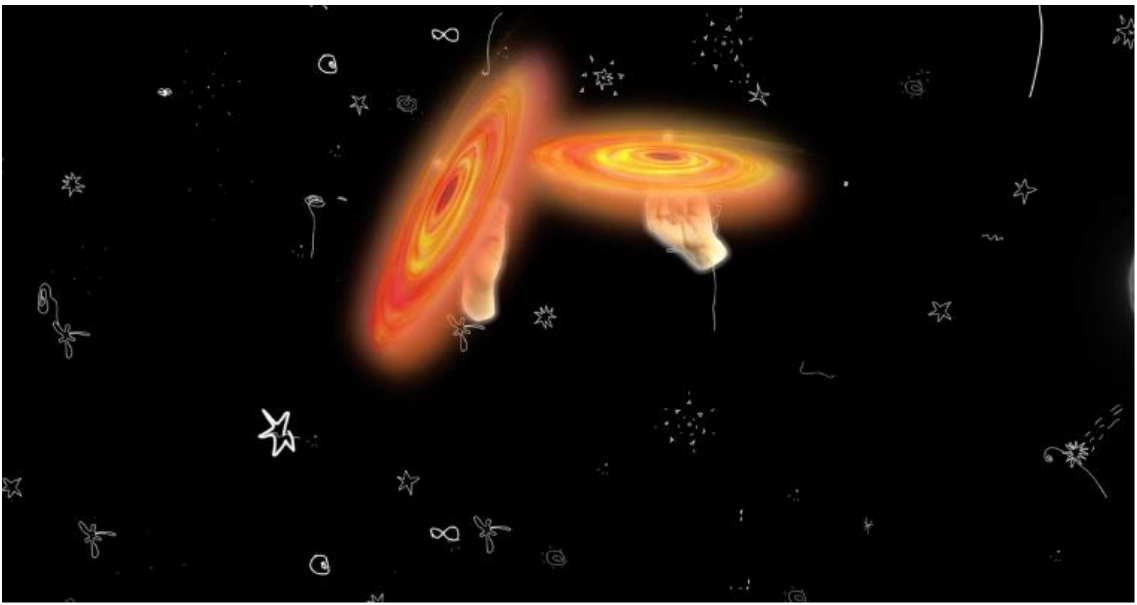

Fig. 6.    The black hole dance. The animation builds on the scientist's hands to convey the notion of dance.1

**Case Study 2: Dark Matter and Galaxy Formation**

In contrast, Scientist #2 used her hands more frequently and dramatically, influenced by her cultural background. In *The Movements of the Universe*, when discussing how dark matter holds galaxies together, she mimicked the act of holding and squeezing a galaxy. This gesture inspired a playful animation where her hands control the galaxy's movements, set against abstract, fairy-tale-like visuals. Her feedback emphasised how gestures helped her intuitively carry both scientific meaning and emotional expression, showing that gestures can convey both emotion and scientific content and serve as powerful tools for communication. It was a way to liberate energy, be more expressive, and to connect with the audience.

---

[1]  https://vimeo.com/839895776/5a6d6cb5bb

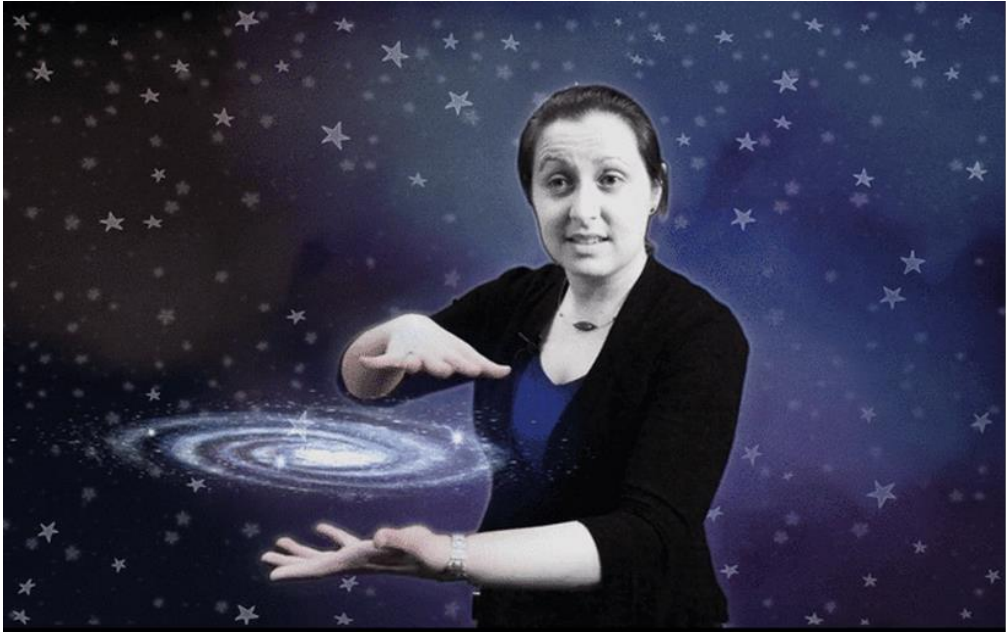

Fig 7. Scientist controlling a galaxy's movements with her hands.2

These case studies reveal how hand gestures, which are often disregarded in scientific communication, can be crucial in making complex astronomical concepts more relatable and engaging. My goal in integrating these gestures into animations is to convert abstract scientific ideas into visual concepts that encourage further exploration and curiosity.

**Astro-Animation with Scientists and Artists**

After exploring how animation can visualise abstract astronomical concepts in *The Movements of the Universe*, we decided to apply these techniques in a series of public workshops to reach broader audiences. My science collaborator and I launched a series of Astro-Animation workshops, bringing together both scientists and artists. These workshops served as a testbed for engaging new audiences, specifically youth from underrepresented communities. Participants from both fields engaged with the process, further highlighting the potential of animation as a tool for scientific communication This collaborative effort between art and science enriched the project and helped establish a foundation for future public workshops.

**Participatory Astro-Animation Experience**

A key aim of the Astro-Animation project is to engage broader audiences—particularly teenagers with limited exposure to science education—by making complex astronomical concepts accessible through a creative, hands-on approach. (Bullock, 2017). These workshops demystify scientific concepts, making them more approachable and fostering curiosity in both science and art. The workshops are continuously evolving as we gather feedback and refine our methods to better engage diverse audiences.

Key areas of focus include evaluating whether the workshops spark curiosity and engagement, determining if participants can creatively express and communicate complex astronomical ideas through animation, and assessing whether the workshops make astronomy more approachable, especially for those who may feel intimidated by science. Ultimately, we aim to see if this creative process breaks down barriers to understanding and participation in science.

---

2  https://vimeo.com/manage/videos/543315132/45fc2d5c10

Each session begins with a brief, accessible lecture by an astronomer who introduces a particular astronomical phenomenon—such as Pluto or Solar Eclipses—followed by a short lecture on the principles of animation. Participants then engage in the hands-on animation phase. To gather insights, we will use surveys and interviews. Participants complete short surveys at the beginning and the end of the workshop. These surveys will measure changes in interest, understanding, and attitudes toward science and art over the course of the workshop. Additionally, we are conducting brief interviews with the participants during the workshop to capture their experiences and reflections

**Astro-Animation Workshops: Public Engagement in Practice**

The Astro-Animation workshops, initially developed in collaboration with scientists, have since been adapted for broader public engagement, including local communities and teenagers. By using animation to simplify complex astronomical concepts, these workshops inspire curiosity and foster interest in both science and art. They have become a key tool for making scientific learning more accessible and engaging for all participants.

**The New York DOT Astronomy Conference**

Building on the success of these workshops, I proposed an Astro-Animation workshop at the New York City DOT Astronomy Conference[3] during the unconference phase, which garnered significant interest and was approved to proceed. The primary goal of this workshop was to engage both scientists and astronomy enthusiasts in a novel method of visualising complex astronomical phenomena, using animation to demystify concepts such as NASA's New Horizons flyby of Pluto (Olkin et al., 2017). The unconference format—flexible and participant-driven—proved to be an ideal environment for this type of collaborative, hands-on experimentation, fostering open dialogue and creative exploration.

This workshop was particularly significant as it invited attendees, many of whom had little to no prior experience with animation, to actively participate in the creative process. By combining scientific insights with artistic techniques, participants were encouraged to explore astronomical phenomena through both visual and imaginative lenses. This hands-on experience not only allowed them to engage more deeply with the subject matter but also reinforced one of the core goals of the Astro-Animation project: making astronomy more accessible and engaging through artistic expression.

**Workshop Structure and Execution**

The session began with a brief, accessible introduction to NASA's New Horizons mission and its flyby of Pluto, which served as the scientific foundation for the animation activity. Participants were then given the opportunity to create their own animated sequences, using printed images from the flyby as a starting point. By encouraging participants to stylise and enhance these frames with their own imagination, the workshop emphasised the intersection of science and art, demonstrating how creative approaches can demystify abstract and complex scientific concepts.

Incorporating animation into educational settings has been shown to engage participants emotionally while also enhancing their cognitive understanding of complex topics (Schwan & Buder, 2015). This approach promotes "active learning", where participants construct knowledge through hands-on processes, such as animation creation. As a result, scientific topics like astronomy become more accessible and engaging to a diverse audience, making it easier for participants to grasp and retain complex ideas (Miller & Fahy, 2009).

**Post-Workshop Film Creation**

On the final day of the conference, I gathered the participants' drawings and shot them using an IPEVO camera. The footage was then imported into Premiere for editing, where I compiled the animations into a short film. This final film, showcasing the participants' work, was presented to the conference attendees on the last day. Watching their drawings come to life through animation was a thrilling and rewarding experience for many participants. It not only provided a sense of accomplishment but also allowed them to see how their individual contributions could be transformed into a cohesive visual narrative, further reinforcing the workshop's goal of merging science with creative expression.

---

[3] https://www.dotastronomy.com/twelve

**Survey and Reflection**

To assess the effectiveness of the workshop, participants were invited to complete a survey at the end of the session. Many of the responses indicated that the activity offered a unique and enjoyable way to engage with scientific concepts. Some of the key feedback included:
- "Very fun! As someone who loves space and art, this is an amazing idea to get people more involved and engaged in astronomy."
- "Colouring was very therapeutic, but I kept feeling like I was going to do it wrong…"
- "It was fun and engaging – looking forward to the results!"

This feedback not only validated the workshop's approach but also provided insights into how participants engaged with science through a creative, hands-on process. For many, the act of creating animated sequences allowed them to explore scientific concepts in a new and accessible way, underscoring the importance of integrating artistic expression into science communication. Furthermore, the therapeutic effect of the activity, as mentioned by several participants, suggests that animation can serve as both an educational and emotional tool for engagement.

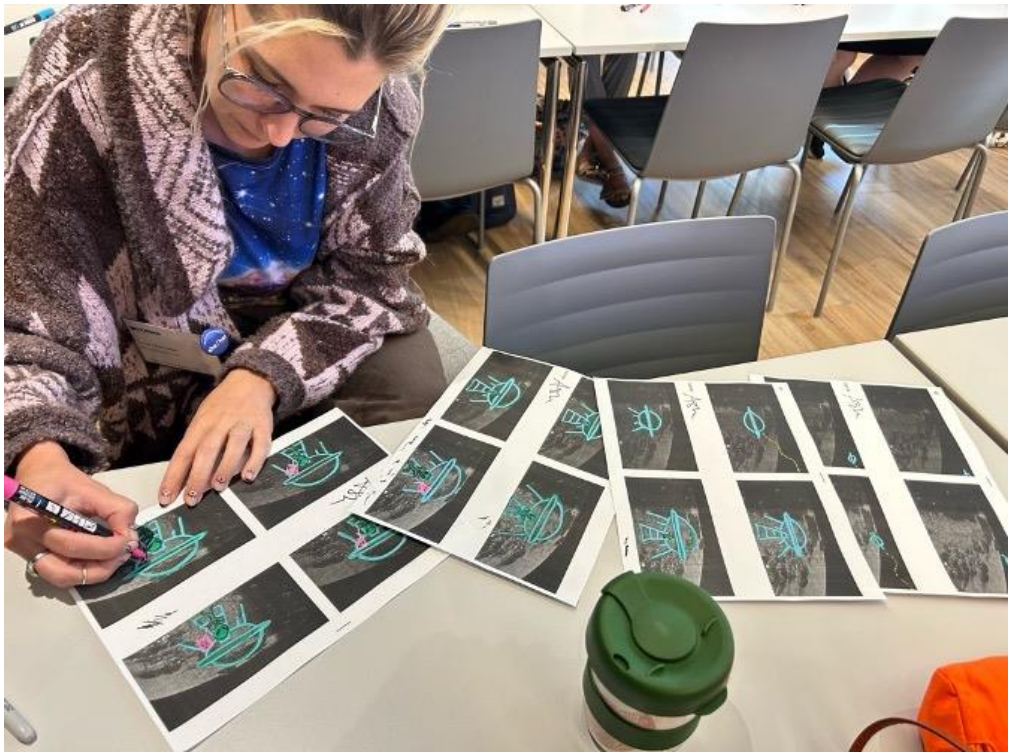

**Fig 8.** Participant during an astro-animation workshop at the Dot.Astronomy conference.

**Workshop Takeaways and Improvements**

The DOT Astronomy workshop provided valuable insights and served as a blueprint for refining the Astro-Animation method. Several key takeaways emerged:

- **Clearer Explanation of Animation Principles** – While participants enjoyed the activity, some felt uncertain about the basic principles of animation. To address this, future workshops now include a brief introductory segment on animation fundamentals, ensuring that all participants feel comfortable and confident in the process. Providing more detailed guides on animation techniques and clearer instructions during the workshop helps participants feel more secure in their artistic choices.

- **Improved Technical Setup** – Capturing the drawings proved time-consuming and could be streamlined by scanning the images instead of shooting them with a camera. This change will result in a faster turnaround and higher quality.

- **Adding participants' interviews** – I have received feedback indicating that many participants have interesting stories to share about their drawings and their thoughts on Pluto. Some of the drawings feature abstract geometrical shapes, while others are purely fantastical. For example, one participant drew a flying saucer with an alien emerging from it, showcasing her creativity. As a result, I have decided to conduct interviews with the participants to have them explain their drawings.

With these improvements, subsequent workshops—such as one conducted at the American Astronomical Society meeting in Texas — were even more successful. During the solar eclipse, participants created animations that were both scientifically informed and artistically engaging. The resulting film was selected for two festivals. Based on these results, more workshops will be conducted in the future, with a particular focus on engaging teenagers.

**Broader Impact: Expanding Public Engagement**

The Astro-Animation project serves as both a creative and educational tool, providing a unique way to connect the public with astronomy. By focusing on underserved teenagers and communities, the project aims to make astronomy more accessible, exciting, and approachable. Hands-on workshops, such as those conducted at the DOT Astronomy Conference, continue to be refined to ensure they effectively engage diverse audiences.

From an artist's perspective, this project highlights the power of visual storytelling in communicating complex scientific ideas. By integrating artistic techniques, such as metaphor, abstraction, and imaginative interpretation, audiences can engage with science on an emotional level. This creative approach makes astronomy more tangible and relatable to participants. Astro-Animation bridges the gap between the factual and the emotional, sparking curiosity and wonder in ways that traditional methods may not.

By merging art and science, Astro-Animation not only brings the universe to life but also fosters a greater interest in STEM fields, offering participants an opportunity to explore the wonders of the universe in an imaginative and meaningful way.

**4.3 How an artist and animation student perceives the universe**

As this essay has posited, artists and scientists have very different views of the universe. Scientists can chart the universe and, from data, can visualise the workings of a vast and complicated universe. Conversely, I believe that artists see the universe and its beauty and, in their own way, try to seek pattern and narrative in its chaos. They are two somewhat opposite ways of seeing, but are not necessarily opposing. Scientists and artists may perceive the universe differently, but they are both trying to see. They all look up at the sky to see and categorise and relate to what they find.

As an artist, I have never focused on science all that much. In school, I was always fairly interested in learning about science - was even quite good at it - but when the time came to focus more on my artistic education, I was admittedly quick to drop it. That is, until a couple of years ago, when I was introduced to a world in which I could utilise my skills in art and animation to communicate and play with science: Astro-Animation.

I was introduced to the world of Astro-Animation as a university student, where I took the aforementioned Astro-Animation class- a class that fuses scientific education and the process of animation. I was able to participate in the class that my fellow authors worked so hard to bring to life and see firsthand how the worlds of science and art can interact. I have since graduated with a BFA in Animation with a Sequential Art minor and can safely say that my experiences in this class shaped the trajectory of the rest of my college career.

Astro-Animation is ultimately an attempt to connect and unify the views of the artist and the views of the scientist and it was my first foray back into working with science. It has not only helped me connect to science in a way that is creatively fulfilling, but has also helped shape my art in many ways going forward. Since I was introduced to Astro-Animation, the throughline of my work has been finding the emotional core behind scientific initiatives and utilising science as a narrative tool.

**Astro-Animation**

**Applying art and narrative to research**

In the spring semester of 2023, I had the privilege of taking the Astro-Animation class. As described above, it was a course where I and a small team of fellow animation students would work with a NASA scientist and create an animated short based on the current project they were engaged with. In this instance, our scientist was working on a project concerning the Lunar Flashlight- a satellite, now decommissioned, meant to orbit and scan the lunar south pole for traces of exposed water ice (Cohen et al., 2020).

Our job was to take that premise and create just about anything we wanted with it. But how does one take such a specific and complicated scientific topic and make it artistic? How do the animators create something that both they and the scientists can be excited about? Will the science have to be compromised for the sake of the art or vice versa? These are all questions that the Astro-Animation class evokes and my team and I handled it a few ways.

**Understanding the reason the research is being done**

The first step that I and my teammates took to creating our animation was to try and understand why the Lunar Flashlight mission existed to begin with. We could easily understand the mechanics of the mission and its scanning of the lunar surface, but we needed to understand the reason that research was being done. Why it was created and what purpose that research was going to serve moving forward were important context clues to start shaping our animation.

In essence, the purpose of the mission was to scan the lunar south pole craters for traces of ice. Any data recovered therein would determine the most appropriate places to land and recover samples from in future missions. Eventually, these samples of ancient ice deep in the southern craters of the Moon, undisturbed by the Sun or anything else for that matter, could uncover information from potentially as far back as the formation of the Earth itself. And that- that little nugget of curiosity and undisturbed secrets- is what artists like myself can latch onto when creating an animated short.

**Translating research into narrative**

The next step in taking broad scientific research and creating something that reflects both parties was finding a narrative that conveyed the mission accurately enough without fully devolving into a simple documentary style exposition. I consider myself a storyteller above anything else and so creating an emotional throughline in our piece was a priority- and habit- of mine.

What we ended up with was a small story about a crew of scientists (who were simply personified versions of the four lasers used to scan the lunar surface) as they search the southern crater. In the end, they happen upon a figure trapped in ice who, once released, tells the crew the story of how life on Earth came to be. In doing so, our team was able to use the characters and the structure of the story not just to showcase what the Lunar Flashlight was for, but as more of a metaphor for the true purpose of its mission.

**The results of bridging art and science**

So, how does one take science and turn it into animation in a way that both honours the work of the NASA scientists and also engages the creativity and passion of the animators? Can it be done? In my case, my team and I found the most success when trying to compromise between the science and the art. We didn't want to fully do away with our narrative tendencies, so instead we used them to enhance and highlight the actual motivation behind the Lunar Flashlight project.

I believe that this approach was the most successful that our group saw and it was one that satisfied both parties in the end. As a group, we enjoyed bringing a story to life that we loved and that also honoured the scientists we were working with. In turn, we saw the scientists feel much less precious about the particulars of their work (which was not the case for every team) because they felt assured that their intentions were being accurately reflected.

The Astro-Animation class was a great study in working across fields and was my first real step into using art and narrative as a tool to convey science to an audience. By focusing on the intentions behind the Lunar Flashlight mission as an emotional throughline for the final animation, me and my classmates were able to handle the balance of art and science in a way that successfully honoured both.

**NASA Internship/Astro-Animation Exhibition**

**Making science and animation accessible**

During the summer of 2023, my time was spent as an intern at the NASA Goddard Space Flight Center. I worked in part as graphics support for Goddard and the other half of the time I worked directly to assist with the Astro-Animation exhibition development. My job was to ideate and prototype different hands-on activities related to animation that could potentially be used for the exhibition. However, animation is complicated and time-consuming work, so the issue became finding activities that could be short enough, simple enough, and fun enough for all ages.

What types of activities could fit these criteria and even then, would they be effective in also teaching science on top of that? This was the main question I based all of my work around moving forward.

**Creating hands-on activities for the exhibition**

Short, simple, and fun was my guiding principle as I set out to find the right activities for the exhibition. My first steps were to make a list of each project I came up with, then log the materials required to make them, and finally prototype each option and time how long it would take to create. By doing so, I could weed out activities that were either too complicated or too lengthy to complete. By the end, there was actually a very interesting trend to the remaining activities.

All of the most successful prototypes I created stemmed from very early forms of animation, such as old optical illusions and crafts. Activities like flipbooks, zoetropes, and phenakistoscopes, well known milestones in the development of animation, all fit the qualifiers of short, simple, and fun projects. This made them perfect candidates for the exhibition.

**Accessibility of animation**

It made sense that all of the most successful prototypes were the ones based on well-established forms of animation because those are all the ones that came about before our modern techniques and technologies. The reason for this is a matter of accessibility. Early animation was very limited in its scope, and so in turn was very accessible in terms of its materials and its simplicity. They were essentially toys and optical illusions, which made them perfect activities for people of all ages and perfect for the activities I was seeking.

Almost all of the participants of the Astro-Animation workshop have never tried animating before and so finding accessible activities was crucial, not just for the enjoyment of the participants, but also to not detract from the scientific topics being discussed in the workshop. After all, the exhibition was meant to foster a dialogue between art and science. Either one becoming too complicated would pull focus away from the true purpose of the exhibition- bringing art and science together.

**The results of creating Astro-Animation activities**

By the end of my internship, I found that the main issue I faced when creating activities for the exhibition all came down to the matter of accessibility. Was the animation accessible? Was the science? All of these things were important to balance and I believe that I eventually found a group of activities that fit those criteria.

Science and animation are both very complex processes in their own right, and it was very important to me that these activities were the perfect balance of engaging and educating. An exhibition like this one can only succeed as long as the participants are immersed and interested, so by simplifying the kinds of art being made I could create the best atmosphere possible, one where the science and the art uplift each other and create one unified experience.

**Senior Thesis Project**

**Using science as narrative**

The culmination of all that I learned from engaging with science through animation was in the creation of my senior thesis project. For a whole year, from the fall semester of 2023 to the spring semester of 2024, I had to dedicate myself to a fully self-produced animated short. I knew from the start that I wanted to create a piece that utilised the knowledge I had gained from working with the concepts of Astro-Animation, however I wanted to take it in a completely new direction than before. But what other ways could science and art be employed?

My work in the Astro-Animation class had been focused on translating science into art and my work for the exhibition had been focused on making science and art accessible to all audiences. All of this served well for the kind of work being done, but a thesis project isn't necessarily meant to teach or even necessarily meant to be accessible. It is meant only to be a reflection of the animator's interests and goals, a reflection of their soul and the culmination of all they have learned up to that point. And so for my thesis project I decided to experiment with using science not as an inspiration for narrative, but as narrative itself.

**Thesis Project**

"*Twin Suns*", my four-and-a-half-minute animated thesis, is about a young woman's journey through feelings of grief at the loss of her sister as well as her isolation in the depths of space. This piece not only utilises space as its setting, but also as a visual metaphor for the main character's emotional journey throughout the film. Through the use of colour and texture, the character's arc is paralleled in the environment around her.

For example, in the film, the use of the colour blue is meant to represent the main character and so, in the first half of the piece, she is often surrounded by blue and other cool, muted colours. Orange is shown to represent the twin character, a ghost that haunts her, and so we see orange and other warm colours start to seep into the woman's world as she learns to accept and live with her grief.

The science that is used along with these colour themes is related to the climax of the animation, where the main character comes across two suns orbiting each other until they finally collide and unite into a brilliant supernova. The suns parallel the relationship between the main character and the twin whom they carry around with them. By coming to terms with the loss of her twin and the acceptance of this grief, so too do the suns unite and create something beautiful.

**The results of using science as narrative**

When I set out to create my thesis project, I wanted to make a piece that honoured my time working with Astro-Animation and I believe I achieved that and so much more. I was able to expand upon all I had learned about science and narrative and create a whole new way of utilising them in my work, one that combines the two. As a narratively minded artist, this very much played to my strengths. I was able to create a story that was meaningful to me and use science as a way to enhance and broaden out that story.

This merging of narrative and science was not only a personal victory for me, but a critical success as well. In October of 2024, *Twin Suns* won for Best Animated Short at the Subtropic Film Festival in Palm Beach, Florida. This was both an incredible honour and also a prime example of how animation can create an environment that makes science accessible and emotionally moving to broader audiences.

I believe that this version of working with science and animation was a resounding success, however that doesn't mean it is necessarily a universal thing. Science as narrative is simply another clever tool to add to the toolbelt of anyone who wishes to explore science and art. In this instance, narrative science just so happened to perfectly fit the needs of myself and my project and so that is what I came up with. However, I do believe that this way of exploring narrative, along with the other ways of working with science and animation, can all be successful if used in their own appropriate contexts.

**Closing Thoughts to A Student's View**

Since I was introduced to Astro-Animation, the throughline of my work has been finding the emotional core behind scientific initiatives and utilising science as a narrative tool. As a student and as an artist, working with science has been a fulfilling pursuit and has been a way to inspire and enhance narrative in my work. It has shifted the way I see the world and has been a way to reconnect with scientific concepts through an artistic lens.

I have also learned a lot about the ways that science and art can be connected to make them accessible to an audience. Over my many projects, my work with the Astro-Animation exhibition has shown me that science and art can blend together in a way that is narratively satisfying and also in a way that makes them both more accessible to a broad audience. By creating an atmosphere of fun and learning, I found that Astro-Animation can help simplify and humanise what are otherwise rather complex pursuits.

Artists and scientists have different ways of seeing the universe- this we know. Yet we are connected in so many ways. Art and science both see and categorise and relate and tell stories and much, much more. Astro-Animation has been a great way of finding those connections. By creating a way to so closely work with these two fields as a combined experience, the exhibition offers a view of science and art that is not often seen- one that is both uplifting, approachable, and altogether unique.

## 5. Educational and Public Engagement Outcomes

Astro-animation has been remarkably successful across both academic and public settings. The Astro-animation class has generated over 70 short films, screened annually at NASA during Take Your Child to Work Day before large and enthusiastic audiences. Every year, more scientists express interest in collaborating with student animators, often taking the resulting films to scientific conferences and using them in their own outreach. Students have also been offered summer internships at NASA; they have invited to co-present at conferences and have contributed to published papers.

The Astro-animation workshops have also proven highly effective, reaching a wide range of participants from scientists to families and teenagers. Evaluation combined surveys, interviews, and observations. Key findings include:
- **Increased Engagement**: Participants reported greater curiosity about astronomy after participating, especially when animation was tied to storytelling and metaphor.
- **Emotional Resonance**: Teenagers used animation to express personal identity and emotional responses to space.

- **Learning Gains**: Many participants could articulate new scientific concepts post-workshop, often grounded in visual or embodied understanding.
- **Toolkit Adaptability**: Feedback led to clearer animation instruction, improved technical workflow, and better facilitation strategies.

Data triangulation revealed that while survey scores alone showed modest change, qualitative interviews highlighted deeper cognitive and affective engagement, especially among teens and first-time animators.

## 6. Discussion

Astro-animation reframes science communication as an embodied, collaborative, and emotionally resonant process. By leveraging metaphor, gesture, and participatory design, it transforms abstract astrophysical ideas into accessible and memorable experiences.

The approach challenges the boundary between scientific fact and creative interpretation, demonstrating that narrative and aesthetics can coexist with scientific rigor. It also foregrounds inclusion by offering a multimodal entry point into astronomy—particularly impactful for underrepresented groups with limited access to STEM.

These findings suggest several broader implications for the fields of science communication and interdisciplinary education. First, they invite a reconsideration of how science is taught in creative education settings, advocating for approaches that value emotional engagement and narrative exploration alongside factual instruction. Second, they point to the potential of interdisciplinary toolkits—such as the Astro-animation model—for use in informal learning contexts, including libraries, festivals, and public workshops. Finally, they highlight the importance of recognizing emotion, gesture, and metaphor as legitimate and powerful forms of knowledge-making. These elements not only support comprehension but also foster a deeper, more personal connection to scientific content.

## 7 Overall Conclusion

Astro-Animation has demonstrated the potential of blending art and science to engage the public with complex astronomical ideas in innovative ways. By fostering collaboration between artists, scientists, and students, this project has opened new avenues for science communication, making abstract concepts more approachable and relatable to broader audiences. This interdisciplinary approach not only enriches public understanding but also provides new creative methods for scientific outreach.

The success of Astro-Animation lies in its ability to demystify science through storytelling, turning data into visual experiences that resonate with audiences on both intellectual and emotional levels. As this project continues to evolve, it offers exciting possibilities for engaging new generations in science education, particularly by making STEM fields more inclusive and appealing to diverse communities.

Looking ahead, the integration of artistic expression with scientific research will likely inspire further projects that encourage exploration, curiosity, and creative learning. Astro-Animation exemplifies how collaboration between these fields can inspire not only understanding of the cosmos but also a deeper connection to the world around us. By continuing to build on these foundations, we can expect to see art and science working together to transform public engagement with science in new and meaningful ways.

**Acknowledgments:**

We thank the many animators, scientists and others who have contributed to this project since its inception. We also thank the two anonymous reviewers for their useful comments. This work was made possible by grants from the Zaentz foundation, the Maryland Space Business Roundtable, the National Endowment for the Arts, and was supported in part by NASA under award number 80GSFC24M0006.

**References**

Arcadias, L. (2021). *The Movements of the Universe – Research* [Video, password: movements]. Vimeo. https://vimeo.com/839895776

Arcadias, L. (2021). *The Movements of the Universe – Research (alternative version)* [Video, password: movements]. Vimeo. https://vimeo.com/543315132

Arcadias, L. & Corbet, R. H. D., An Astro-Animation Class: Optimizing Artistic, Educational, and Outreach Outcomes. *Leonardo* (2022) 55 (4): 414–420. https://doi.org/10.1162/leon_a_02241

Arcadias, L., Corbet, R. H. D., McKenna, D., & Potenziani, I. (2020). Astro-animation: A case study of art and science education. *Animation Practice, Process & Production*, *9*(1), 75–102. https://doi.org/10.1386/ap3_000018_1

Archer, L., Dawson, E., DeWitt, J., Seakins, A., & Wong, B. (2015). "Science capital": A conceptual, methodological, and empirical argument for extending Bourdieusian notions of capital in science education research. *Journal of Research in Science Teaching*, *52*(7), 922–948. https://doi.org/10.1002/tea.21227

Atwood, W. B., Abdo, A. A., Ackermann, M., et al. (2009). The Large Area Telescope on the Fermi Gamma-Ray Space Telescope mission. *The Astrophysical Journal*, *697*(2), 1071–1102. https://doi.org/10.1088/0004-637X/697/2/1071

Avraamidou, L., & Osborne, J. (2009). The role of narrative in communicating science. *International Journal of Science Education*, *31*(12), 1683–1707. https://doi.org/10.1080/09500690802380695

Bruner, J. S. (1986). *Actual minds, possible worlds*. Harvard University Press.

Bullock, E. C. (2017). Only STEM can save us? Examining race, place, and STEM education as property. *Educational Studies*, *53*(6), 628–641. https://doi.org/10.1080/00131946.2017.1369082

Cohen, B. A., Hayne, P. O., Greenhagen, B., Paige, D. A., Seybold, C., & Baker, J. (2020). Lunar flashlight: Illuminating the lunar south pole. *IEEE Aerospace and Electronic Systems Magazine*, *35*(3), 46–52. https://doi.org/10.1109/MAES.2020.2969997

Crafton, D. (1991). *The hand of the artist in early animation*. In K. Thompson (Ed.), *Hollywood animation: The emergence of a genre* (pp. 29–57). Routledge.Crary, J. (1990). *Techniques of the observer: On vision and modernity in the nineteenth century*. MIT Press.

Egan, K. (1986). *Teaching as storytelling: An alternative approach to teaching and curriculum in the elementary school*. Althouse Press.

Funk, S. (2015). Ground- and space-based gamma-ray astronomy. *Annual Review of Nuclear and Particle Science*, *65*, 245–277. https://doi.org/10.1146/annurev-nucl-102014-021957

Hamza, A. (2024). The intersection of art and science: A multidisciplinary perspective. *Kashf Journal of Linguistics*.

Kendrew, S., Simpson, R. J., Lintott, C. J., Crawford, S. M., Smith, A., Ödman-Govender, C., Bauer, A. E., Smethurst, R., & Nekoto, W. (n.d.). Ten years of astronomy: Scientific and cultural impact. Retrieved October 1, 2024, from https://en.wikipedia.org/wiki/Foo_Camp

Miller, S., Fahy, D., & ESConet Team. (2009). Can science communication workshops train scientists for reflexive public engagement? The ESConet experience. *Science Communication*, *31*(1), 116–126. https://doi.org/10.1177/1075547009339048

Olkin, C. B., Ennico, K., & Spencer, J. (2017). The Pluto system after the New Horizons flyby. *Nature Astronomy*, *1*(10), 663–670. https://doi.org/10.1038/s41550-017-0228-x

Roberts, M. S. E. (2013). Surrounded by spiders! New black widows and redbacks in the Galactic field. *Proceedings of the International Astronomical Union*, *291*, 127–132. https://doi.org/10.1017/S1743921312023983

Roth, W.-M., & Lawless, D. V. (2002). When up is down and down is up: Body orientation, proximity, and gestures as resources for language and learning in science. *Science Education*, *86*(3), 355–377. https://doi.org/10.1002/sce.10009

Salimpour, S., Bartlett, S., Fitzgerald, M. T., et al. (2021). The gateway science: A review of astronomy in the OECD school curricula, including China and South Africa. *Research in Science Education*, *51*, 975–996. https://doi.org/10.1007/s11165-020-09922-0

Shapiro, S. L., & Teukolsky, S. A. (1983). *Black holes, white dwarfs, and neutron stars: The physics of compact objects*. Wiley-Interscience.

The Electromagnetic Spectrum | HubbleSite. (n.d.). Retrieved October 1, 2024, from https://hubblesite.org/contents/articles/the-electromagnetic-spectrum